\documentclass[letter]{aa}

\usepackage{newtxtext,newtxmath}

\usepackage[T1]{fontenc}
\usepackage{ae,aecompl}

\usepackage{natbib}
\bibpunct{(}{)}{;}{a}{}{,} 

\usepackage{graphicx}   
\usepackage{amsmath}    
\usepackage{amssymb}    
\usepackage{xcolor}
\usepackage[normalem]{ulem}
\usepackage{subcaption}
\usepackage{booktabs}

\providecommand{\gaia}{\textsl{Gaia }}

\providecommand{\msunnospace}{$M_\odot$}

\newcommand{\kmskpc}{km$~$s$^{-1}~$kpc$^{-1}$ }
\newcommand{\kmskpcnospace}{km$~$s$^{-1}~$kpc$^{-1}$}

\providecommand{\lmc}{$\mathnormal{G}_{\text{LMC}}$ }
\providecommand{\smc}{$\mathnormal{G}_{\text{SMC}}$ }
\providecommand{\mw}{$\mathnormal{G}_{\text{MW}}$ }
\providecommand{\lmcnospace}{$\mathnormal{G}_{\text{LMC}}$}
\providecommand{\smcnospace}{$\mathnormal{G}_{\text{SMC}}$}
\providecommand{\mwnospace}{$\mathnormal{G}_{\text{MW}}$}

\begin{document}

\title{Tidal interaction can stop galactic bars: \\ on the LMC non-rotating bar}

\author{Ó. Jiménez-Arranz\inst{1}
   \and S. Roca-Fàbrega\inst{1}
}

\institute{{$^1$ Lund Observatory, Division of Astrophysics, Department of Physics, Lund University, Box 43, SE-22100, Lund, Sweden}}

\date{Received <date> / Accepted <date>}

\abstract 
{Using \gaia DR3 data, \citet{jimenez-arranz24a} computed the LMC bar pattern speed using three different methods. One of them suggested that the LMC might be hosting a bar that barely rotates, and is slightly counter-rotating with respect to the LMC disc, with a pattern speed of $\Omega_p = -1.0 \pm 0.5 $ \kmskpcnospace.}
{To confirm that tidal interactions can trigger the LMC hosting a non-rotating bar due to its interaction with the SMC, which could cause the LMC bar to slow down significantly until it (momentarily) stops.} 
{We analyse a subset of models (K9 and K21) from the KRATOS suite \citep{jimenez-arranz24b} where we detected non-rotating bars. We make use of two different methods to track the evolution of the bar pattern speed: the program \texttt{patternSpeed.py} \citep{dehnen23}, and temporal finite-differences of the change rate the bar major axis' phase angle.} 
{In the second LMC-SMC-like pericenter passage of K9, the bar of the LMC-like galaxy weakens to almost disappear and regenerates with a pattern speed that suffers a slowdown from $\Omega_p \sim 20$ \kmskpc to $\Omega_p \sim 0$ \kmskpc in less than 75 Myr. Then, the bar rotates at less than $\Omega_p \sim 3-5$ \kmskpc for around 100 Myr, until it recovers the initial (before interaction) pattern speed of $\Omega_p \sim 10$ \kmskpcnospace. The results for the K21 simulation are comparable.}
{This work is the first direct evidence that galactic bars can be slowed down or even stopped by tidal interaction, which strengthens the possibility of the LMC hosting a non-rotating bar, and can add an alternative formation scenario for observed slow-rotating bars.}



\maketitle

\section{Introduction}
\label{sec:introduction}

Galactic bars are ubiquitous. About two-thirds of spiral galaxies in the local Universe host them \citep[e.g.][]{eskridge00,masters11,erwin18}. Because bars can serve as both sources and sinks of angular momentum, redistributing stars, gas, and dark matter within galaxies, they are believed to be key drivers of secular evolution \citep[e.g.][]{athanassoula02,athanassoula03,debattista06,sellwood14}. Galactic bars rotate almost rigidly, whose rotation is parametrised by its angular frequency (or pattern speed), and are formed through global disc instabilities. In isolation, this instability can be produced by the transfer of angular momentum from the bar to the dark matter halo \citep[e.g.][]{sellwood80,weinberg85,sellwood14}. Simulations show that the bar pattern speed gradually decreases over time as a result of dynamical friction \citep[e.g.][]{sellwood80,athanassoula03,dehnen23,jimenez-arranz24b}.

Simulations have also shown that the tidal interaction of a disc galaxy with a massive companion is another way of boosting global disc instabilities and, consequently, the formation of galactic bars \citep[e.g.][]{gerin90,lokas14,ghosh21,jimenez-arranz24b,ansar25,zheng25}. In interaction, the evolution of the barred galaxy is then controlled by the subsequent pericenter passages, and the effects will depend on the orientation of the bar with respect to the tidal torque from the host at the pericenter. Some studies have shown that bars can be destroyed (and later regenerated) by the interaction between galaxies \citep[e.g.][]{lang14,cavanagh22,jimenez-arranz24b,ansar25,zheng25}. In a recent paper, we have also shown that the tidal torque can speed up or slow down the bar, changing its pattern speed and strength \citep{jimenez-arranz24b}.

From an observational point of view, the analysis of the LMC, that is the closest barred galaxy to the Milky Way (MW), allows for detailed studies with unprecedented resolution, providing key insights into the formation and evolution of barred galaxies in general \citep[e.g.][]{luri20,Niederhofer2022,jimenez-arranz23a,kacharov24,jimenez-arranz25}. The LMC's bar is quite peculiar, since it is off-centred and tilted with respect to the LMC plane \citep[e.g.][]{vdmcioni01,Choi2018,luri20,rathore25,jimenez-arranz25}. Simulations suggest that this uncommon setup may be produced by the recent interaction with one of its satellite galaxies, the SMC \citep[e.g.][]{besla12,jimenez-arranz24b}, which happened around 150-200 Myr ago \citep[e.g.][]{diaz-bekki12}.

Recently, \citet{jimenez-arranz24a} determined the LMC bar pattern speed using \gaia DR3 data \citep{gaiadr3summary}. Surprisingly, one of the methods employed suggested that the LMC might be hosting a bar that barely rotates, and is slightly counter-rotating with respect to the LMC disc, with a pattern speed of $\Omega_p = -1.0 \pm 0.5 $ \kmskpcnospace. 

Our goal in this work is to evaluate whether the tidal interaction between the LMC and the SMC could cause the LMC bar to abruptly slow down until it (momentarily) stops. We test this scenario using suitable numerical simulations, such as the KRATOS suite of simulations \citep{jimenez-arranz24b} of LMC-SMC-MW-like galaxies.

\section{KRATOS simulations}
\label{sec:simulations}

In the main body of this work we analyse the K9 simulation of the KRATOS suite \citep{jimenez-arranz24b}\footnote{However, it is worth mention that we observed slow or almost stopped bars in other interacting simulations in KRATOS (see Sect. \ref{sec:discussion}), showing that this may not be a particular case for a particular configuration or stability of the LMC-like stellar disc, but that requires of future analysis.}. KRATOS consists of a suite of 28 pure N-body simulations that model the evolution of an LMC-like galaxy with varying parameters. The suite includes models where the LMC-like galaxy is in isolation, in interaction with an SMC-mass galaxy, or in interaction with both an SMC-mass galaxy and a MW-mass galaxy. In this work, we use the notation presented in \citet{jimenez-arranz24b}, where the LMC-like galaxy is denoted as \lmcnospace, and the SMC- and MW-mass systems as \smc and \mwnospace, respectively. K9 is one of the KRATOS suite simulations that includes all three galaxies (\lmcnospace, \smcnospace, and \mwnospace). The simulation used in this paper is briefly described in the following paragraphs; therefore, for all the details, we recommend that the reader consult the primary reference \citep{jimenez-arranz24b}.

In K9, as in all simulations of the KRATOS suite, the \lmc system is modelled as a stellar exponential disc embedded in a live dark matter Navarro-Frenk-White \citep[NFW, ][]{nfw96} halo. The \smc system is modelled as a simple NFW halo. Both \smc dark matter and stellar particles are generated at once following the NFW profile. However, in this work, as in \citet{jimenez-arranz24b}, the particles with the strongest gravitational binding are later defined as the stellar component of the SMC for analysis and visualization purposes. As all particles, DM and stellar, are treated as collisionless point-like sources of gravity, we emphasize that this selection procedure has no effect on the models. We simply sought to capture the evolution of the stellar component and its interaction with the surrounding environment by using this particle selection strategy. Lastly, we only model the MW DM content in \mwnospace, ignoring the MW disc's contribution to the total mass of \mwnospace. This is because our primary focus is on the effects that the interaction between the three galaxies produces in the \lmc disc.

The temporal and spatial resolutions of the simulations are 5000 years and 10 pc, respectively. Each particle has a minimum mass of $4 \times 10^3$\msunnospace. All simulations have been run for 4.68 Gyr, starting at the apocenter of the LMC-SMC's second interaction. Originally, every KRATOS suite simulation is made up of 61 snapshots ($\delta t \sim 78$ Myr). Some models, however, have been re-run to save snapshots with a higher temporal cadence, increasing the number of snapshots per simulation to 2,003 ($\delta t \sim 2$ Myr)\footnote{With 61 snapshots per simulation, the original KRATOS suite is open-data and available online at \url{https://dataverse.csuc.cat/dataset.xhtml?persistentId=doi:10.34810/data1156}. On reasonable request, the high temporal cadence version (2,003 snapshots) of the KRATOS suite can be provided.}. In this work, we employ the high cadence version of the K9 simulation. Having a high temporal cadence enables us both to have a detailed evolution of the \lmc bar pattern speed and to determine the bar pattern speed with finite-differences with high precision (see Sect. \ref{sec:methodology}). Additionally, the snapshot cadence is sufficiently high to examine the bar structure's dynamics. In fact, the Nyquist theorem \citep{shannon49} guarantees that we can track and recover frequencies up to 0.245 \kmskpcnospace.

Finally, our study focuses on the second \lmcnospace-\smc pericenter passage\footnote{Which takes place at $t = -0.136$ Gyr, following the decision made in \citet{jimenez-arranz24b} where the snapshot taken after 4.0 Gyr from the initial conditions is considered to be $t=0$. For the discussion of this decision, we direct the reader to the primary reference.} and has an impact parameter of 1.68 kpc; that is, the \smc is crossing the \lmc disc. In K9, the \lmc has a bar formed with a pattern speed $\Omega_p$ of $\sim 10$ \kmskpc prior to the second pericenter passage. Then, this bar weakens to almost disappear due to the \smc interaction and recombines to form a new bar. For a more detailed description of the \lmcnospace-\smc interaction and the \lmcnospace's bar properties, we refer the reader to Sects. 3 and 6 of \citet{jimenez-arranz24b}, respectively. The \mw is at least $\gtrsim 150$kpc from the \lmcnospace-\smc system during the simulation time analysed in this paper. Therefore, its impact on the internal kinematics of the \lmcnospace, and specifically its bar pattern speed, is minimal.

\section{Measurement of the bar pattern speed}
\label{sec:methodology}

To examine the bar features of the \lmc disc, we use the same centring and alignment process as in \citet{jimenez-arranz24b}. After that, we make use of two different methods to track the evolution of the bar pattern speed, which provide robustness to our analysis and results. First, we use the program \texttt{patternSpeed.py} \citep{dehnen23}. This method is an unbiased, precise, and consistent method that simultaneously measures the bar pattern speed $\Omega_p$ and the orientation angle $\phi_b$ of the bar from single snapshots of simulated barred galaxies. These parameters are found assuming that: $1)$ the continuity equation applies; $2)$ the centre of rotation is known; $3)$ that the rotation is around the $z'$-axis, and; $4)$ that the density is stationary in the rotating frame. The pattern speed of bars has already been estimated using this method extensively in both simulations \citep{bland-hawthorn23,hey23,machado24,jimenez-arranz24b,semczuk24} and 6D data of real galaxies \citep{jimenez-arranz24a,zhang24}.

The program \texttt{patternSpeed.py} determines the pattern speed in the bar region, which is defined by $[R_0,R_1]$. The method allows to either automatically find the bar region by searching for large amplitude of the bisymmetric density perturbation of second order and having a roughly constant phase angle (see their Appendix B for details) or that the user directly inputs bar region. We chose the second option and the bar region was fixed to be $[R_0,R_1] = [0.2, 1.0]$ kpc throughout the entire time evolution. Since the pattern speed is taken to be constant throughout the entire bar, we are confident that, by making this cautious selection on the bar length $R_1$, we get reliable and comparable results between snapshots. On the other hand, an overestimation of the bar length $R_1$ would imply the contamination of different pattern speeds, such as the spiral arms, which could bias our results.

Second, taking advantage of the high temporal cadence of the re-run KRATOS suite of simulations ($\delta t \sim 2$ Myr), we are also able to compute the bar pattern speed by means of finite-differences of the change rate the bar major axis’ phase angle $\phi_b$. We average the rate of change of the phase angle of the bar major axis $\phi_b$ in three consecutive snapshots over the radial range of the bar to determine the bar pattern speed. Also, for this method, and throughout the entire time evolution, we define the same bar region as for the  program \texttt{patternSpeed.py}, which is $[R_0,R_1] = [0.2, 1.0]$ kpc.

\section{Results}
\label{sec:results}

\begin{figure*}[t!]
    \centering
    \includegraphics[width=1\textwidth]{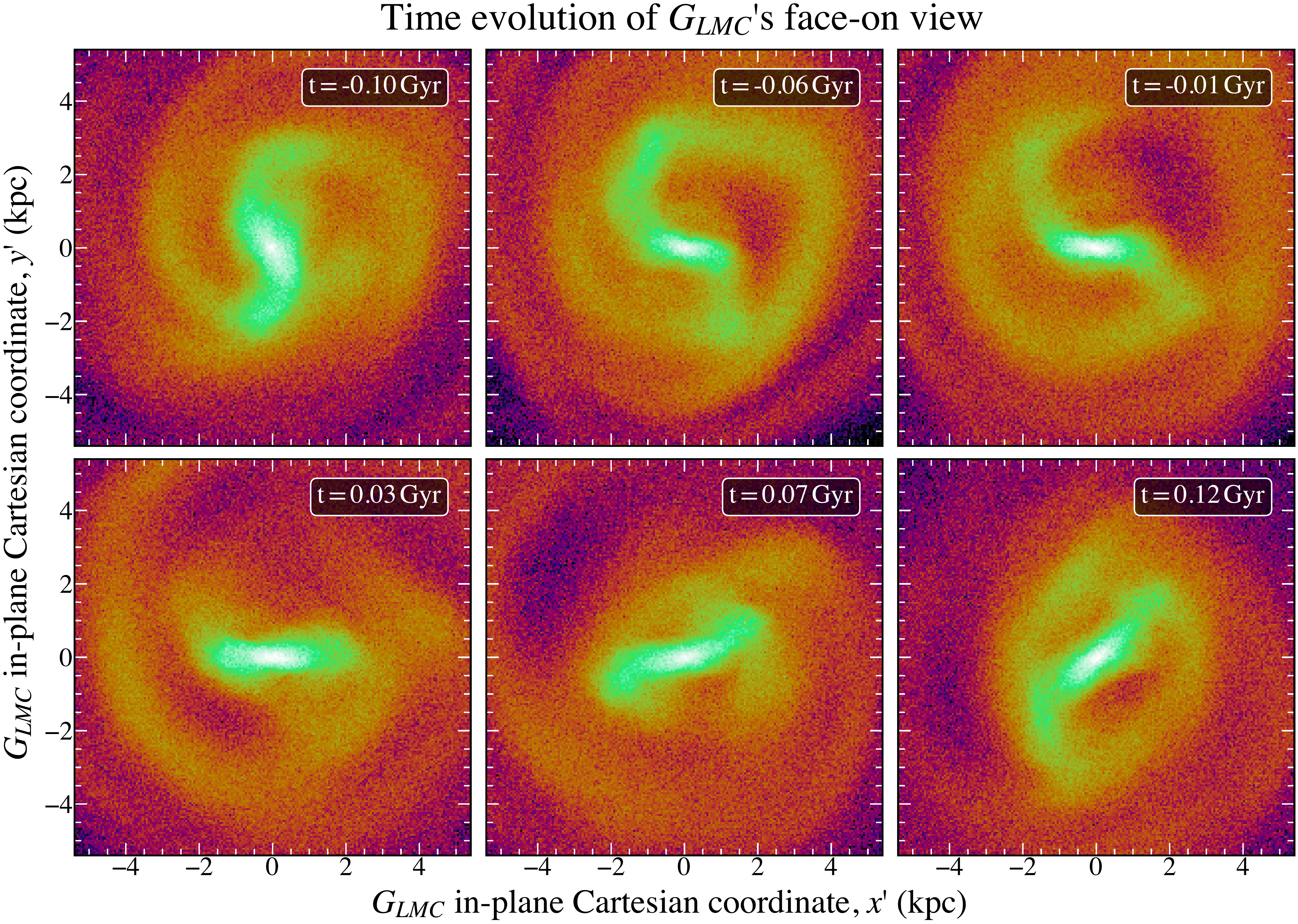}
    \caption{Time evolution of the stellar density map of the \lmc disc as seen face-on. Panels have a constant 45 Myr time difference, with a total time evolution of 220 Myr (from $t=-0.10$ Gyr to $t=0.12$ Gyr). All maps are shown in the \lmc in-plane $(x', y')$ Cartesian coordinate system. A video version of this Figure is available online.}
    \label{fig:macroplot}
\end{figure*}

First, we start with analysing \lmcnospace's bar pattern speed using a qualitative visual approach. Figure \ref{fig:macroplot} shows the evolution of the \lmc disc's stellar density structures\footnote{A video of the time evolution of the stellar density map of the \lmc disc is made available online at \url{https://www.oscarjimenezarranz.com/visualizations/tidal-interaction-can-stop-galactic-bars}.} in a period of 220 Myr, in a face-on view of the disc (see Sect. \ref{sec:methodology} for the definition of the disc plane). In this Figure, we show snapshots that are close to the second pericenter passage between the \lmc and \smc which occurs shortly before the image shown in the first panel, at $t = -0.136$ Gyr in simulation units ($t = 0.0$ Gyr corresponds to the time when the KRATOS fiducial simulation resembles the most the current observations of the \lmcnospace). All the panels are shown from the simulated box reference frame, which is by construction ``inertial'' (no bulk velocity or rotation). We can see that although most structures evolve and rotate around the center, the bar major-axis remains aligned with the $x'$-axis for about 130 Myr (second to fourth panel), which is just after the pericenter of the \smcnospace. Before and after this event (first and last panels), the bar also rotates around the center like all other structures.

\begin{figure}[t!]
    \centering
    \includegraphics[width=\columnwidth]{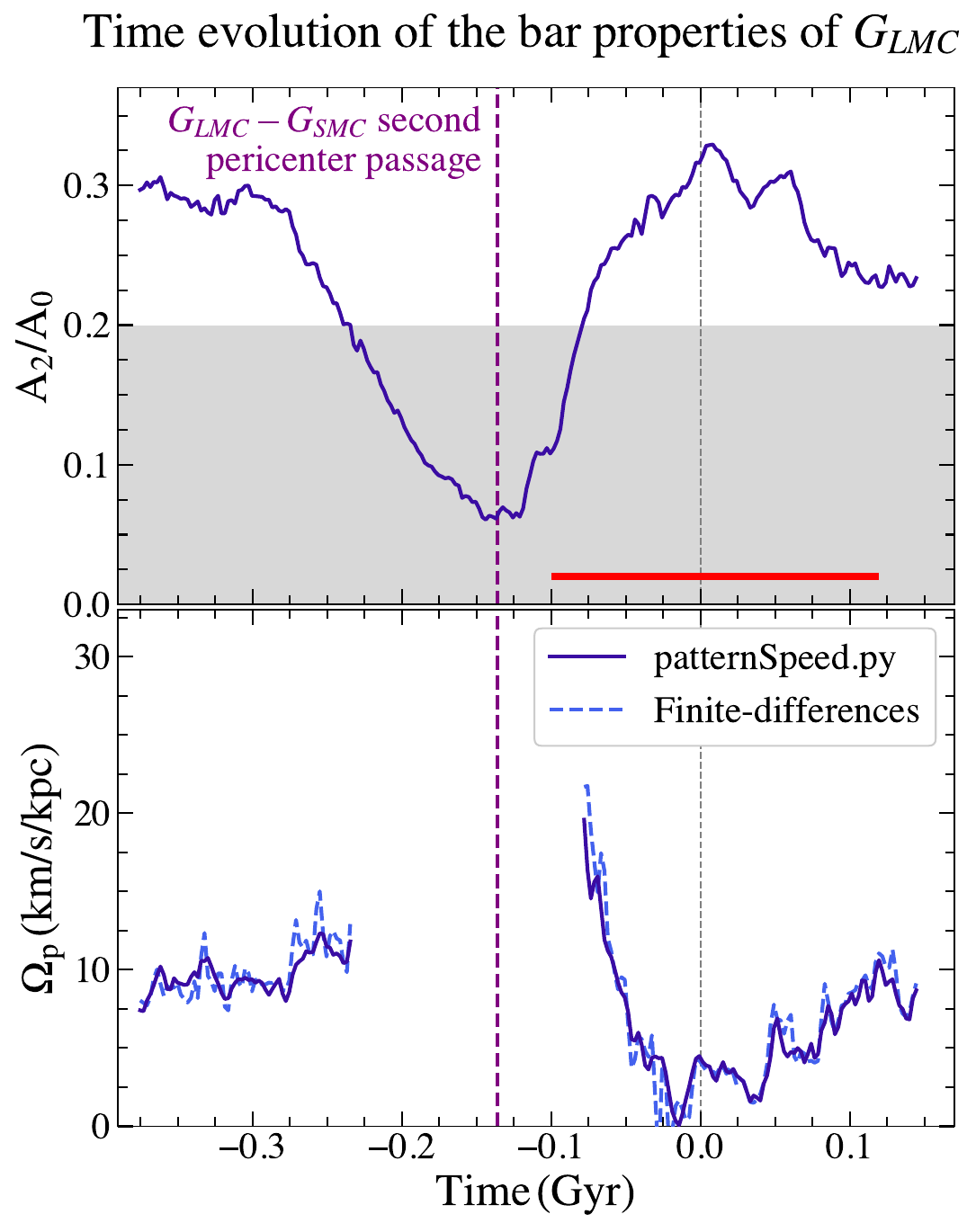}
    \caption{Top: Time evolution of the relative $m=2$ Fourier amplitude $A_2/A_0$ in the \lmcnospace's bar region -- defined by $[R_0,R_1] = [0.2, 1.0]$ kpc. Bottom: Bar pattern speed $\Omega_p$  determined by the program \texttt{patternSpeed.py} (blue solid) and by means finite-differences (blue dashed). The grey area in top panel corresponds to $A_2 / A_0 < 0.2$, which is the threshold used to consider whether or not the \lmc disc has a bar. The horizontal red line corresponds to the time interval shown in Fig. \ref{fig:macroplot} (from $t=-0.10$ Gyr to $t=0.12$ Gyr). In the bottom panel, we only show the bar pattern speed $\Omega_p$ when the \lmc disc has a bar ($A_2 / A_0 > 0.2$). The vertical purple dashed line corresponds to the \lmcnospace-\smc pericentre.}
    \label{fig:omegap_time}
\end{figure}

The former qualitative evaluation requires a quantification using the analysis techniques introduced in Sect. \ref{sec:methodology}. With this analysis, we aim to provide a quantitative analysis of \lmcnospace's bar pattern speed $\Omega_p$ by means of two independent methods: the program \texttt{patternSpeed.py} and by means of finite-differences on the rate of change of the phase angle of the bar major axis $\phi_b$. Figure \ref{fig:omegap_time} shows the time evolution of the relative $m=2$ Fourier amplitude $A_2/A_0$  (top panel) and the bar pattern speed $\Omega_p$ (bottom panel) in the \lmcnospace's bar region -- defined by $[R_0,R_1] = [0.2, 1.0]$ kpc (see Sect. \ref{sec:methodology}). The grey area in top panel corresponds to $A_2 / A_0 < 0.2$, which is the threshold used to consider whether or not the \lmc disc has a bar. In the bottom panel, we only show the bar pattern speed $\Omega_p$ when the \lmc disc has a bar ($A_2 / A_0 > 0.2$). The vertical purple dashed line corresponds to the \lmcnospace-\smc second pericentric passage. We show the bar pattern speed $\Omega_p$ determined by the program \texttt{patternSpeed.py} and by means finite-differences with a blue solid and dashed line, respectively. We notice that both methods provide consistent results. The reader may observe that the time interval shown in Fig. \ref{fig:omegap_time} (from $t=-0.38$ Gyr to $t=0.15$ Gyr) is larger than that of Fig. \ref{fig:macroplot} (from $t=-0.10$ Gyr to $t=0.12$ Gyr, horizontal red line) in order to contextualize what we see in the latter. 

In Fig. \ref{fig:omegap_time}, we can observe how prior the \lmcnospace-\smc interaction, the \lmc has a bar with strength $A_2/A_0 \sim 0.3$ and pattern speed $\Omega_p \sim 10$ \kmskpcnospace. Then, the \lmc gets deprived from a bar around 100 Myr before the \lmcnospace-\smc second pericenter passage, and recovers it around 50 Myr after the interaction. The new \lmcnospace's bar grows into a similar strength than before ($A_2/A_0 \sim 0.25- 0.3$). However, its pattern speed suffers a slowdown from $\Omega_p \sim 20$ \kmskpc to $\Omega_p \sim 0$ \kmskpc in less than 75 Myr. Then, \lmcnospace's bar rotates at less than $\Omega_p \sim 3-5$ \kmskpc for around 100 Myr, until it recovers the initial pattern speed of $\Omega_p \sim 10$ \kmskpcnospace. Remarkably, we have the non-rotating bar near $t=0$ (dashed grey vertical line), which would be consistent with the observations \citep{jimenez-arranz24a}.

\section{Discussion}
\label{sec:discussion}

Using \gaia DR3 data, the authors of \citet{jimenez-arranz24b} used three distinct methods to determine the LMC bar pattern speed. This study worked with a dataset of $\sim$12 million LMC stars with full astrometric information, that had been cleaned from MW foreground contamination \citep{jimenez-arranz23a}. Of these, $\sim$30,000 stars had also line-of-sight velocity information. The quality and richness of \gaia data allowed for the evaluation of the LMC bar pattern speed using three different methods. Those were: the Tremaine-Weinberg \citep[TW,][]{tw84} method, a bisymmetric velocity \citep[BV,][]{drimmel22} model, and the program \texttt{patternSpeed.py} \citep[referred as the ``Dehnen method'',][]{dehnen23}.

Surprisingly, the third method, the program \texttt{patternSpeed.py} suggested that the LMC might be hosting a bar that barely rotates, and is slightly counter-rotating with respect to the LMC disc, with a pattern speed of $\Omega_p = -1.0 \pm 0.5 $ \kmskpcnospace. The viability of this result was discussed in \citet{jimenez-arranz24a}. Examples of situations in which this LMC non-rotating result could be biased and non-physical are; 1) a possible strong and counter-rotating $m=1$ disc component, which could balance the bar pattern speed, and would not be taken into account by the method; 2) the method may be sensitive to dust extinction and completeness effects in the inner LMC region, perhaps more strongly than the other methods. In that work was also suggested, though, that the pattern speed that \texttt{patternSpeed.py} recovers may actually be the real LMC bar pattern speed, and that the bar deceleration and slight counter-rotation could be the result of interaction with the SMC and/or the MW. This paper is one of the natural follow-up projects of \citet{jimenez-arranz24a}, in which we use KRATOS \citep{jimenez-arranz24b}, a suite of numerical simulations of the LMC-SMC-MW system, to assess the viability of the LMC hosting a non-rotating bar due to the interaction with the SMC.

The subject of galaxies with bars that have nearly zero pattern speed has been covered in only a few articles. Some numerical simulations do contain bars with such odd property, but only for extremely particular and challenging configurations. For example, \citet{2023Collier} run a numerical experiment of an $N$-body galaxy embedded in a counter-rotating (retrograde) live dark matter halo which acts as a reservoir of negative angular momentum. A bar embedded in a counter-rotating dark matter halo can decelerate, then flip its pattern speed, and finally decoupling its rotation from the disc. This letter provides an alternative formation scenario for the observed slow-rotating bars \citep[e.g.][]{2009chemin,buttitta22} or even non-rotating bars \citep[e.g.][]{jimenez-arranz24a}, that is a tidal interaction with a companion galaxy.

It is worth mentioning that the main body of this work focuses on the K9 simulation of the KRATOS suite. However, this is not the only \lmcnospace-\smc interacting simulation inside the suite where we can observe slow or almost stopped bars \citep[see Fig. 10 of][]{jimenez-arranz24b}. For instance, K21 is another simulation in which the \lmc bar stops to be nearly non-rotating (see Appendix \ref{sec:appendix}). It differs from K9 in terms of the baryonic and total mass of \lmc and its Toomre $Q$ parameter, owing to a less massive and more stable \lmcnospace. We highlight that the orbits in K21 differ from those in K9 due to the change in total mass, which affects their direct comparability to current orbit estimates. However, this difference provides a valuable opportunity to explore the impact of mass variations on orbital dynamics, making it a worthwhile subject of investigation -- for the details of K21 we refer the reader to \citet{jimenez-arranz24b}. 

The presence of multiple models with stopped bars suggests that this process is not just a unique case in a specific simulation of a particular galaxy. Instead, it may represent a more general phenomenon in the Universe, highlighting the need for further research to understand its broader implications. Attempts have been made to identify and characterize the factors that cause the \lmc bar to stop. Although some efforts have been made to check the radial migration (D. Hebrail, in prep.), tidal torque maps, and the phase space, the primary causes of the extreme bar slowdown remain unknown and elusive.

Overall, in this work we provide, for the first time, direct evidence that tidal interaction can stop galactic bars. More specifically, this work provides support to the idea that the LMC may be hosting a non-rotating bar, produced by the most recent interaction with the SMC, which happened $\sim$ 150-200 Myr \citep[e.g.][]{diaz-bekki12}. However, a more thorough project is needed to provide a detailed characterization and theoretical framework for this phenomenon, which is not the goal of this work.

\section{Conclusions}
\label{sec:conclusions}

In this work we analyse the K9 (and K21, see Appendix \ref{sec:appendix}) simulation of the KRATOS suite \citep{jimenez-arranz24b} to assess the viability of the LMC hosting a non-rotating bar \citep{jimenez-arranz24a} due to the interaction with the SMC. KRATOS consists of a suite of 28 simulations that model the evolution of the LMC-SMC-MW system. 

Our study focuses on the LMC-SMC most recent interaction, which happened around 150-200 Myr ago \citep[e.g.][]{diaz-bekki12}. In K9, the \lmc has a bar that weakens to almost disappear by the interaction with the \smc and then recombines to form a new bar. After the interaction, the pattern speed of this newly formed bar suffers a slowdown from $\Omega_p \sim 20$ \kmskpc to $\Omega_p \sim 0$ \kmskpc in less than 75 Myr. Then, \lmcnospace's bar rotates at less than $\Omega_p \sim 3-5$ \kmskpc for around 100 Myr, until it recovers the initial pattern speed of $\Omega_p \sim 10$ \kmskpcnospace. 

To the best of our knowledge, this work is the first direct evidence that galactic bars can be stopped by tidal interaction. This result strengthens the possibility of the LMC hosting a non-rotating bar \citep{jimenez-arranz24a}, and can add an alternative formation scenario for observed slow-rotating bars \citep[e.g.][]{fragkoudi21}. 

Finally, we would like to emphasize that the purpose of this work is to report that tidal interactions can stop rotating galactic bars and not to provide an in-detail characterisation or theoretical frame to this phenomena, which is left for a more detailed project\footnote{If other research groups are interested in carrying on with this work, we emphasize that the simulation(s) used in this paper are available upon request.}.

\begin{acknowledgements}

OJA acknowledges funding from ``Swedish National Space Agency 2023-00154 David Hobbs The GaiaNIR Mission'' and ``Swedish National Space Agency 2023-00137 David Hobbs The Extended Gaia Mission''. SRF acknowledge financial support from the Spanish Ministry of Science and Innovation through the research grant PID2021-123417OB-I00, funded by MCIN/AEI/10.13039/501100011033/FEDER, EU.

\end{acknowledgements}

\bibliographystyle{aa}
\bibliography{mylmcbib} 

\begin{appendix}

\section{K21 simulation}
\label{sec:appendix}

\begin{figure*}[t!]
    \centering
    \includegraphics[width=1\textwidth]{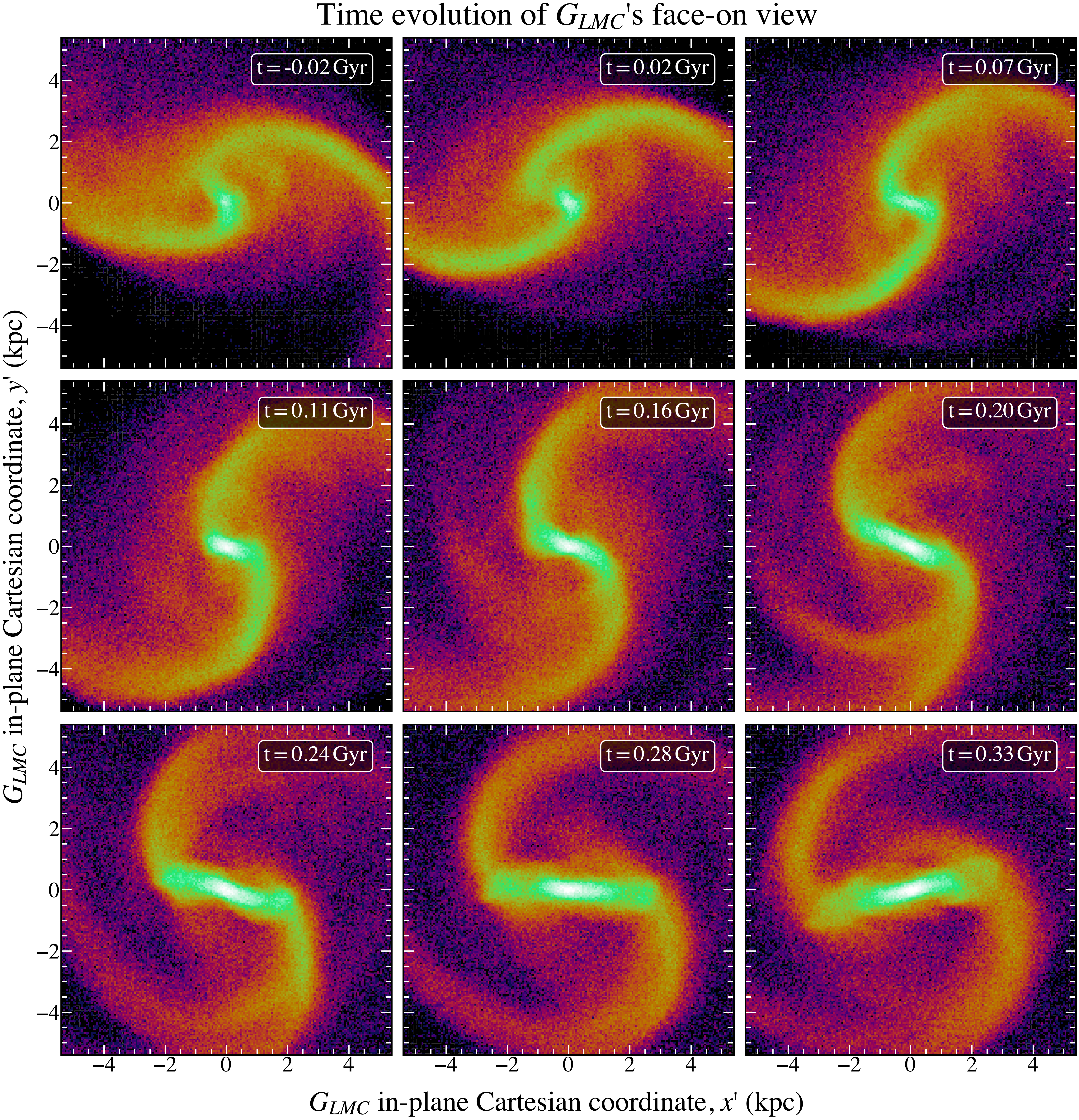}
    \caption{Same as Fig. \ref{fig:macroplot} but for the K21 simulation. Panels have a constant 44 Myr time difference, with a total time evolution of 350 Myr (from $t=-0.02$ Gyr to $t=0.33$ Gyr). A video version of this Figure is available online.}
    \label{fig:macroplot_K21}
\end{figure*}

The results of the evolution of the K21's bar pattern speed over time are displayed in this Appendix. Figure \ref{fig:macroplot_K21} shows the evolution of the \lmc disc's stellar density structures in a period of 350 Myr, in a face-on view\footnote{A video of the time evolution of the stellar density map of the \lmc disc is made available online at \url{https://www.oscarjimenezarranz.com/visualizations/tidal-interaction-can-stop-galactic-bars}.} (see Sect. \ref{sec:methodology} for the definition of the disc plane). In this Figure, we show snapshots that are after the first pericenter passage between the \lmc and \smc -- notice that the orbits in K21 differ from those in K9 due to the change in total mass --  which occurs at $t = -0.303$ Gyr, and far away from the \lmc disc, at a minimum distance of $d = 12.9$ kpc. The arrangement of Fig. \ref{fig:macroplot_K21} is similar to that of Fig. \ref{fig:macroplot}, this time showing 9 panels. All the panels are shown from the simulated box reference frame, which is by construction “inertial” (no bulk velocity or rotation). This illustrates how, for roughly 170 Myr, the bar major-axis stays nearly aligned with the $x'$-axis, despite the fact that most structures change and revolve around the center (third to seventh panel). Like all other structures, the bar rotates around the center before and after this event (the two first and last panels).

\begin{figure}[t!]
    \centering
    \includegraphics[width=\columnwidth]{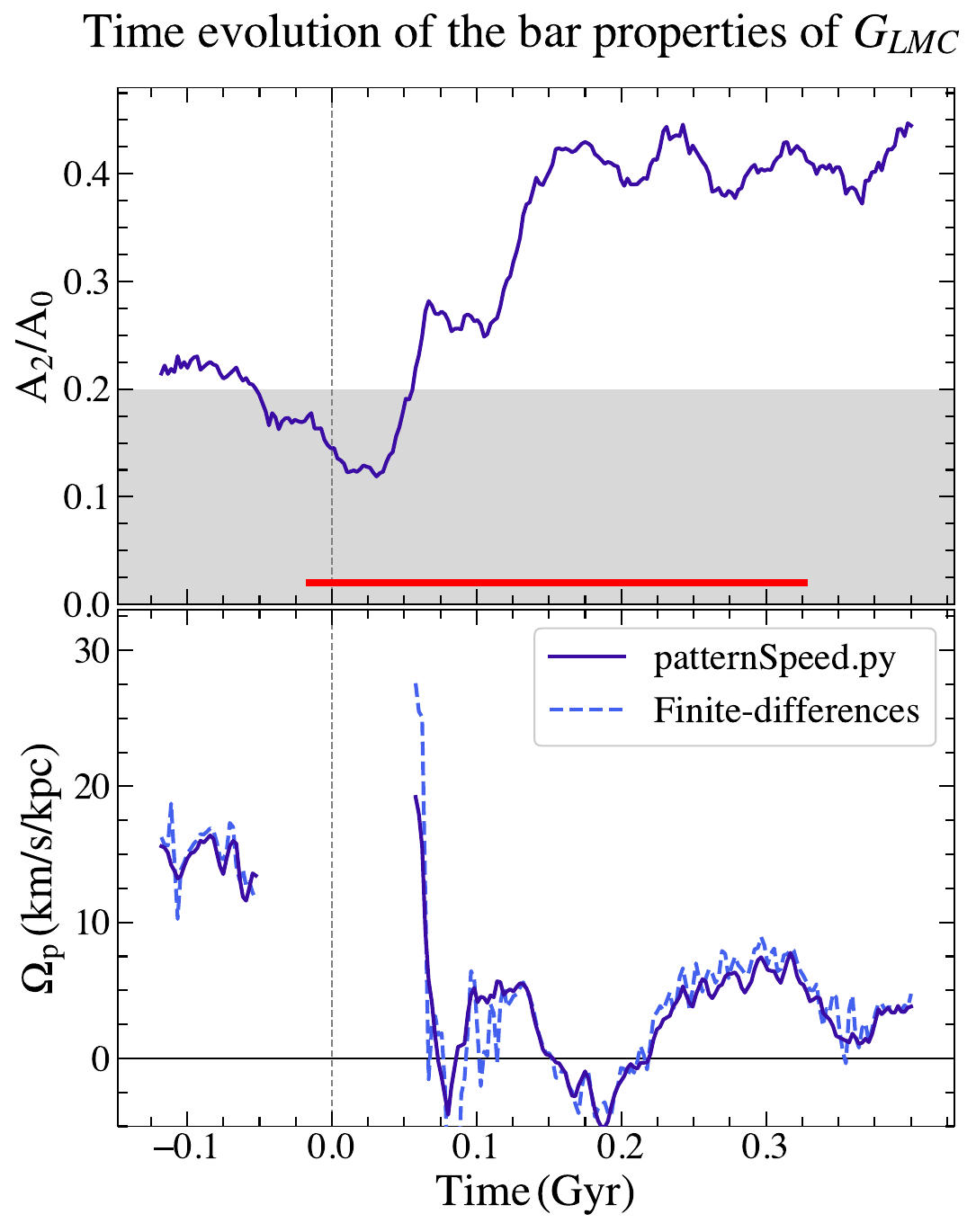}
    \caption{Same as Fig. \ref{fig:omegap_time} but for the K21 simulation. In the top panel, the horizontal red line corresponds to the time interval shown in Fig. \ref{fig:macroplot_K21} (from $t=-0.02$ Gyr to $t=0.3$ Gyr). Outside of the displayed temporal range, at $t = -0.303$ Gyr, the first (and previous) pericenter passage between the \lmc and \smc takes place.}
    \label{fig:omegap_time_K21}
\end{figure}

Figure \ref{fig:omegap_time_K21} shows the time evolution of the relative $m=2$ Fourier amplitude $A_2/A_0$  (top panel) and the bar pattern speed $\Omega_p$ (bottom panel) in the K21 \lmcnospace's bar region -- defined by $[R_0,R_1] = [0.2, 1.0]$ kpc (see Sect. \ref{sec:methodology}). Fig. \ref{fig:omegap_time_K21} follows the same layout as Fig. \ref{fig:omegap_time}. We can observe that prior $t=0$, there is a weak bar ($A_2/A_0 \gtrsim 0.2$) with a roughly constant pattern speed around $\Omega_p \sim 15$ \kmskpcnospace. The bar then becomes weaker and reconvenes at $t=0.05$ Gyr, experiencing a severe slowdown in pattern speed to $\Omega_p \sim 0$ \kmskpcnospace, and for a while even to negative values. After reaching $\Omega_p \sim 5$ \kmskpcnospace, it subsequently slows down to negative bar pattern speed values, peaking at $A_2/A_0 \sim 0.4$ for bar strength.  Finally, the bar recovers a pattern speed of $\Omega_p \sim 5$ \kmskpcnospace.

\end{appendix}

\label{lastpage}

\end{document}